\documentclass[print,showpacs,superscriptaddress,a4paper,10pt,twocolumn,pra]{revtex4}
\usepackage{amsmath}
\usepackage{amsfonts}
\usepackage{amssymb}
\usepackage{graphicx}

\begin{document}

\title{Integrability of two-component nonautonomous nonlinear
Schr\"odinger equation}
\author{Cai-Ying Ding}
\affiliation{Center of Interdisciplinary Studies, Lanzhou University, Lanzhou 730000,
China}
\author{Fang Zhang}
\affiliation{Center of Interdisciplinary Studies, Lanzhou University, Lanzhou 730000,
China}
\author{Dun Zhao}
\affiliation{School of Mathematics and Statistics, Lanzhou University, Lanzhou 730000,
China}
\affiliation{Center of Interdisciplinary Studies, Lanzhou University, Lanzhou 730000,
China}
\author{Hong-Gang Luo}
\affiliation{Center of Interdisciplinary Studies, Lanzhou
University, Lanzhou 730000, China}
\affiliation{Key Laboratory for Magnetism and Magnetic Materials of the Ministry of Education,
Lanzhou University, Lanzhou 730000, China}
\author{W. -M. Liu}
\affiliation{Beijing National Laboratory for Condensed Matter Physics, Institute of
Physics, CAS, Beijing 100190, China}

\begin{abstract}
We investigate the integrability of generalized nonautonomous nonlinear
Schr\"odinger (NLS) equations governing the dynamics of the single- and
double-component Bose-Einstein condensates (BECs). The integrability
conditions obtained indicate that the existence of the nonautonomous soliton
is due to the balance between the different competition features: the
kinetic energy (dispersion) versus the harmonic external potential applied
and the dispersion versus the nonlinearity. In the double-component case, it
includes all possible different combinations between the dispersion and
nonlinearity involving intra- and inter-interactions. This result shows that
the nonautonomous soliton has the same physical origin as the canonical one,
which clarifies the nature of the nonautonomous soliton. Finally, we also
discuss the dynamics of two-component BEC by controlling the relevant
experimental parameters.
\end{abstract}

\maketitle
\section{Introduction}
The soliton is ubiquitous in nature and its existence is due to the
balance of two physical features, the dispersion and the
nonlinearity in a nonlinear system, as observed by Zabusky and
Kruskal \cite{ZK1965}. The particle-like nature of the soliton has
attracted much attention in the last decades since its potential
application in optical soliton communication \cite{HK1995,HW1996}
and its rich dynamics shown in different systems. Particularly, the
realization of the Bose-Einstein condensate (BEC) in 1995
\cite{Anderson1995,Davis1995} provided an ideal ground to study the
dynamics of the soliton by using the Feshbach resonance
\cite{Inouye1998, Kevrekidis2003}, a technique to change the
scattering length of atoms, and by tuning the harmonic external
potential applied. This system can be well described by the
so-called nonautonomous nonlinear Schr\"odinger (NLS) equation with
the harmonic external potential. The dynamics of the soliton
governed by this equation shows quite complicated behavior in
comparison with its canonical counterpart \cite{ZK1965}. While some
early studies \cite{Chen1976,Calogero1976,Konotop1993} have already
broken the concept of the canonical soliton, the study of exact
soliton-like solutions of nonautonomous NLS equation and discussion
of their nature have attracted much attention in recent years \cite
{Serkin2000,Kruglov2003,Serkin2004,liang2005,Serkin2007,Beitia2008,He2009,Kundu2009,Serkin2010}.

After the systematic ways to find the exact soliton-like solutions
of the nonautonomous NLS equation have been proposed
\cite{Serkin2000, Kruglov2003}, the attention has been focused on
the nature of the solutions obtained. In 2003, Kruglov et al. argued
that the self-similar solutions they obtained were not soliton
solutions because of the non-integrability of the equation. However,
under certain conditions the nonautonomous NLS equation is
integrable \cite{Serkin2004, Serkin2007, He2009, Kundu2009,
Serkin2010}. Nevertheless, these soliton-like solutions are quite
different from the canonical ones, they move with varying amplitudes
and speeds. Thus Serkin, Hasegawa and Belyaeva \cite{Serkin2007}
proposed the concept of nonautonomous soliton. Thus a question
arises naturally: Could the nonautonomous and canonical solitons be
viewed in a unified way?

In this paper we indicate that the unified view is actually the
concept of balance between different features. First of all, we
consider the single-component Gross-Pitaevskii (GP) equation with
the time-dependent dispersion, nonlinearity and harmonic external
potential. Although the integrability condition of this equation has
been obtained in the literature \cite{Serkin2007,He2009}, its
physical meanings has not been discussed in a deep way. Here we
point out that in the presence of harmonic external potential such
as BEC, the stability of the nonautonomous soliton can be attributed
to two kinds of competitions and their balance. One is the
competition between the nonlinearity and the dispersion, which is
the standard one, and the other is the competition between the
kinetic energy (dispersion) and the harmonic external potential. The
latter one has not been pointed out explicitly in the literature. To
further deepen understanding of the picture, we consider
two-component GP equation, which includes the intra- and
inter-interactions. The procedure obtaining the integrability
condition by using the Painlev\'e analysis \cite{Weiss1983} in this
case is proved to be nontrivial but the result is physically
straightforward, namely, for each component the competition between
the kinetic and potential energies balance the competitions between
all possible combinations of the dispersion and nonlinearity
including intra- and inter-components (see below). This result shows
sufficiently that nonautonomous soliton and its canonical
counterpart can be viewed in a unified way from the point of view of
balance. Finally, based on the integrability condition in
two-component case we explore the dynamics of nonautonomous two
solitons and the spatial separation of binary BECs in both
integrable and non-integrable cases.

This paper is organized as follows. In Section II we reformulate the
integrability condition of single-component nonautonomous NLS case.
In Section III we explore the integrability for the double-component
case. The integrability condition is composed of the combinations of
different competition features, in which the basic form is similar
to the single-component case. In Section IV we consider some
applications in obtaining nonautonomous soliton solution of the
two-component nonautonomous NLS equation and phase separation of
binary BECs. Finally, Section V is devoted to a brief summary.

\section{Integrability condition of single-component case}

Below we first consider the one-dimensional nonautonomous NLS equation
\begin{eqnarray}
&& i\frac{\partial} {\partial t} u(x,t) + \varepsilon f(t)\frac{\partial ^{2}%
} {\partial x^{2}} u(x,t) + \delta\, g(t) \vert u(x,t)\vert ^{2} u(x,t)
\notag \\
&& \hspace{4cm} -\frac{1}{2} V(t) x^{2} u(x,t) = 0,  \label{eq1}
\end{eqnarray}
where $\varepsilon$ and $\delta$ are dispersion and nonlinearity coefficients in the standard NLS equation and $f(t)$ and $g(t)$
are dispersion and nonlinearity managements, respectively. $V(t)$ is the
time-dependent harmonic external potential. Eq. (\ref{eq1}) has been
extensively studied in the literature by various methods like the Lax pair
\cite{Serkin2000,Serkin2007,Khawaja2010} and the similarity transformation
\cite{Kruglov2003}. In Ref. \cite{He2009}, we used the Painlev\'e analysis
to explore the integrability condition of Eq. (\ref{eq1}) and proposed a
systematic way to find the exact solutions of Eq. (\ref{eq1}) directly from
the known solutions of the standard NLS equation. These solutions can be obtained only under the integrability
condition
\begin{equation}
\xi(t) := 2\varepsilon f(t) V(t) +\frac{d^{2}}{dt^{2}}\ln \frac{f(t) }{g(t) }%
-\frac{d}{dt}\ln g(t) \frac{d}{dt}\ln \frac{f(t) }{g(t) } = 0.  \label{eq2}
\end{equation}
Eq. (\ref{eq2}) is proved to be equivalent to that in Refs. \cite{Serkin2007, Serkin2010}. In the absence of the harmonic external potential,
Eq. (\ref{eq2}) is reduced to the balance condition between the dispersion and nonlinearity. In this case, one obtains
\begin{equation}
g(t) = \frac{f(t)}{c_1 + c_2\int f(t^{\prime }) dt^{\prime }},  \label{eq3}
\end{equation}
where $c_1$ and $c_2$ are two constants determined by the initial dispersion
and nonlinearity. If one takes $g(0) = 1$ and $g^{\prime }(0) = 0$, then $c_1 = f(0)$ and $c_2 = f^{\prime }(0)/f(0)$. Eq. (\ref{eq3}) clearly
manifests the balance between the dispersion and nonlinearity. In contrast
to the canonical case both the dispersion and nonlinearity can be
time-dependent but the system is still integrable and the soliton keeps
stable while its shape and velocity change, as shown in Ref. \cite{Serkin2007, He2009}. When the external potential is present, as in the case
of BEC, it can be balanced by the kinetic energy (dispersion term), as shown
in the first term of Eq. (\ref{eq2}). As a result, Eq. (\ref{eq2}) can be
viewed as the competitions between the kinetic energy and external potential
and between the dispersion and nonlinearity. This can be further
demonstrated by considering two-component nonautonomous NLS equation below.

\section{Integrability condition of two-component case}

The balance can be further clarified by considering two-component NLS equation
\begin{eqnarray}
&& \hspace{-0.5cm} i\frac{\partial}{\partial t} u_i(x,t) +
\varepsilon_{i}f_{i}(t)\frac{\partial ^{2}}{\partial x^{2}} u_i(x,t) +
\delta _{ii}g_{ii}(t)\vert u_i(x,t)\vert ^{2}u_i(x,t)  \notag \\
&& \hspace{0.2cm} + \delta _{ij}g_{ij}(t)\vert u_{j}(x,t)\vert ^{2}u_i(x,t)-\frac{1}{2}V_{i}(t) x^{2}u_i(x,t) = 0,  \label{eq4}
\end{eqnarray}
where the subscripts $i = 1, 2$ and $j = 3-i$ denote the two components,
respectively. As in the single component case, $f_{i}(t)$ and $g_{ii}(t)$
represent the time-dependent dispersion and nonlinearity in the
intra-component and $V_{i}(t)$ is the harmonic external potential. Besides
the intra-component interaction, there is also inter-component interaction $g_{ij}(t)$.

To explore the integrability of Eq. (\ref{eq4}) we follow the
Painlev\'e analysis \cite{Weiss1983}. First of all one can expand
$u_i(x,t)$ and their complex conjugate $v_i(x, t) = u^*_i(x, t)$ on
a non-characteristic singularity manifold $\psi(x,t) = x - \phi(t) $
as follows
\begin{eqnarray}
u_i(x, t) = \sum_{k=0}^\infty u_{i, k}(t) \psi(x, t)^{k-1},  \label{eq5a} \\
v_i(x, t) = \sum_{k=0}^\infty v_{i, k}(t) \psi(x, t)^{k-1}.  \label{eq5b}
\end{eqnarray}
Inserting Eqs. (\ref{eq5a}) and (\ref{eq5b}) into Eq. (\ref{eq4}),
performing the standard leading-order analysis and studying the polynomial
of $\phi(t)$, one obtain the compatibility condition for the two-component
NLS equation as follows
\begin{equation}
\chi (t)+(2a_{11}(t)+2a_{22}(t)-a_{12}(t)-a_{21}(t))\left( \frac{d}{dt}\ln
g(t)\right) ^{2}=0,  \label{eq6}
\end{equation}
where $g(t)=\frac{g_{12}(t)g_{21}(t)}{g_{11}(t)g_{22}(t)}$, $a_{11}(t)=\frac{
\delta _{22}g_{22}(t)}{\delta _{12}g_{12}}$ and $a_{12}(t)=\frac{\varepsilon
_{1}}{\varepsilon _{2}}\frac{f_{1}(t)}{f_{2}(t)}.$ $a_{22}(t)$ and $%
a_{21}(t) $ can be obtained by exchanging the indices of $1$ and $2$ in the
expressions of $a_{11}(t)$ and $a_{12}(t)$. $\chi (t)$ is complicated but
there is a simple picture, it consists of the balances between the kinetic
energy and potential energy and the possible pairs of dispersions and
nonlinearities in the two-component system, namely,
\begin{eqnarray}
&& \hspace{-0.8cm}\chi(t) = a_{11}(t)\left(\frac{1}{\lambda g(t)} - 1\right)
\xi_{1, f_1 \leftrightarrow g_{11}}(t) + (1\rightleftharpoons2)  \notag \\
&& \hspace{-0.1cm} + a_{11}(t)(\lambda g(t) - 1) \xi_{1, f_1\leftrightarrow
g_{12}}(t) + (1\rightleftharpoons2)  \notag \\
&& \hspace{-0.1cm} + \left(a_{12}(t) - \frac{\lambda g(t)}{a_{12}(t)}\right)
\xi_{1, f_2\leftrightarrow g_{21}}(t) + (1\rightleftharpoons2)  \notag \\
&& \hspace{-0.1cm} + \left(a_{12}(t) - \frac{1}{a_{12}(t) \lambda g(t)}%
\right) \xi_{1, f_2\leftrightarrow g_{22}}(t) + (1\rightleftharpoons2).
\label{eq7}
\end{eqnarray}
\begin{figure}[tbp]
\includegraphics[width=0.8\columnwidth]{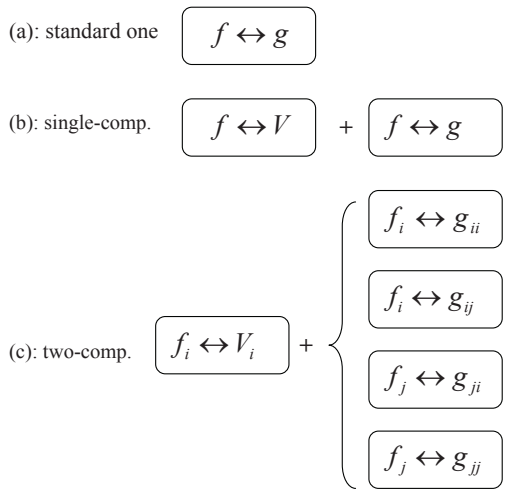}
\caption{The unified physical picture for the balance between the different
competitions for (a) the conventional NLS equation, (b) the single-component
NLS equation and (c) the two-component NLS equation $i=1,2$ $(j=3-i)$ with
harmonic external potential.}
\label{fig1}
\end{figure}
Here the expression of each $(1\rightleftharpoons 2)$ is obtained by
interchanging the subscripts of 1 and 2 of the left term given in each row
correspondingly in the right of the above equation, and $\lambda =\frac{%
\delta _{12}\delta _{21}}{\delta _{11}\delta _{22}}$. Four $\xi $
functions for the component $1$ read
\begin{eqnarray}
&&\hspace{-0.5cm}\xi_{1, f_1\leftrightarrow g_{11}}(t) = 2 \varepsilon_1
f_1(t) V_1 (t) + \frac{d^{2}}{dt^{2}}\ln \frac{f_1(t) }{g_{11}(t)}  \notag \\
&& \hspace{0cm} -\frac{d}{dt}\ln g_{11}(t) \frac{d}{dt}\ln \frac{f_1(t) }{%
g_{11}(t)},  \label{eq7a} \\
&& \hspace{-0.5cm}\xi_{1, f_1\leftrightarrow g_{12}}(t) = 2 \varepsilon_1
f_1(t) V_1 (t) + \frac{d^{2}} {dt^{2}}\ln \frac{g_{22}(t)}{g_{21}(t)}\frac{%
f_1(t)}{g_{12}(t) }  \notag \\
&& \hspace{0cm} -\frac{d}{dt}\ln \frac{g_{21}(t)}{g_{22}(t)}g_{12}(t) \frac{d%
}{dt}\ln \frac{g_{22}(t)}{g_{21}(t)} \frac{f_1(t) }{g_{12}(t)},  \label{eq7b}
\\
&& \hspace{-0.5cm}\xi_{1, f_2\leftrightarrow g_{21}}(t) = 2 \varepsilon_1
f_1(t) V_1 (t) + \frac{d^{2}}{dt^{2}}\ln \frac{f_2(t)}{g_{21}(t) }  \notag \\
&& \hspace{0cm} -\frac{d}{dt}\ln \frac{f_{1}(t)}{f_{2}(t)}g_{21}(t) \frac{d}{%
dt}\ln \frac{f_2(t) }{g_{21}(t)},  \label{eq7c} \\
&&\hspace{-0.5cm} \xi_{1, f_2\leftrightarrow g_{22}}(t) = 2 \varepsilon_1
f_1(t) V_1 (t) + \frac{d^{2}}{dt^{2}}\ln \frac{g_{12}(t)}{g_{11}(t)}\frac{%
f_2(t)}{g_{22}(t) }  \notag \\
&& \hspace{0cm} -\frac{d}{dt}\ln \frac{f_1(t)}{f_2(t)}\frac{g_{11}(t)}{%
g_{12}(t)}g_{22}(t) \frac{d}{dt}\ln \frac{g_{12}(t)}{g_{11}(t)}\frac{f_2(t)}{%
g_{22}(t)}.  \label{eq7d}
\end{eqnarray}
respectively. By the similar way, the $\xi $ functions of the component $2$
can be obtained by inter-exchanging the indices of $1$ and $2$ in the above
four functions. These equations uncover the basic features for the
integrability condition of the two-component NLS equation, as in the
single-component case. That is the balance between the competition between
the kinetic energy and the harmonic external potential and the competition
between the dispersion and nonlinearity. In the two-component case, for each
component there is one pair for the kinetic energy and the harmonic external
potential, which competes with the different pairs of the dispersion and the
nonlinearity, namely, $f_{1}\leftrightarrow g_{11}$, a contribution from the
intra-component; $f_{1}\leftrightarrow g_{12}$, a contribution from the
inter-component; $f_{2}\leftrightarrow g_{21}$ and $f_{2}\leftrightarrow
g_{22}$, due to the existence of the component $2$ and relevant competitions
between the dispersion and the nonlinearity in it. To clarify the above
features, in Fig. \ref{fig1} we show schematically this picture and make
comparison with the cases in the single-component and the standard NLS
equations. This result indicates that the integrability condition in these
equations can be viewed as a unified picture, irrespective of the standard
or nonautonomous systems.

\section{Applications}

\begin{figure}[h]
\includegraphics[width=0.45\columnwidth]{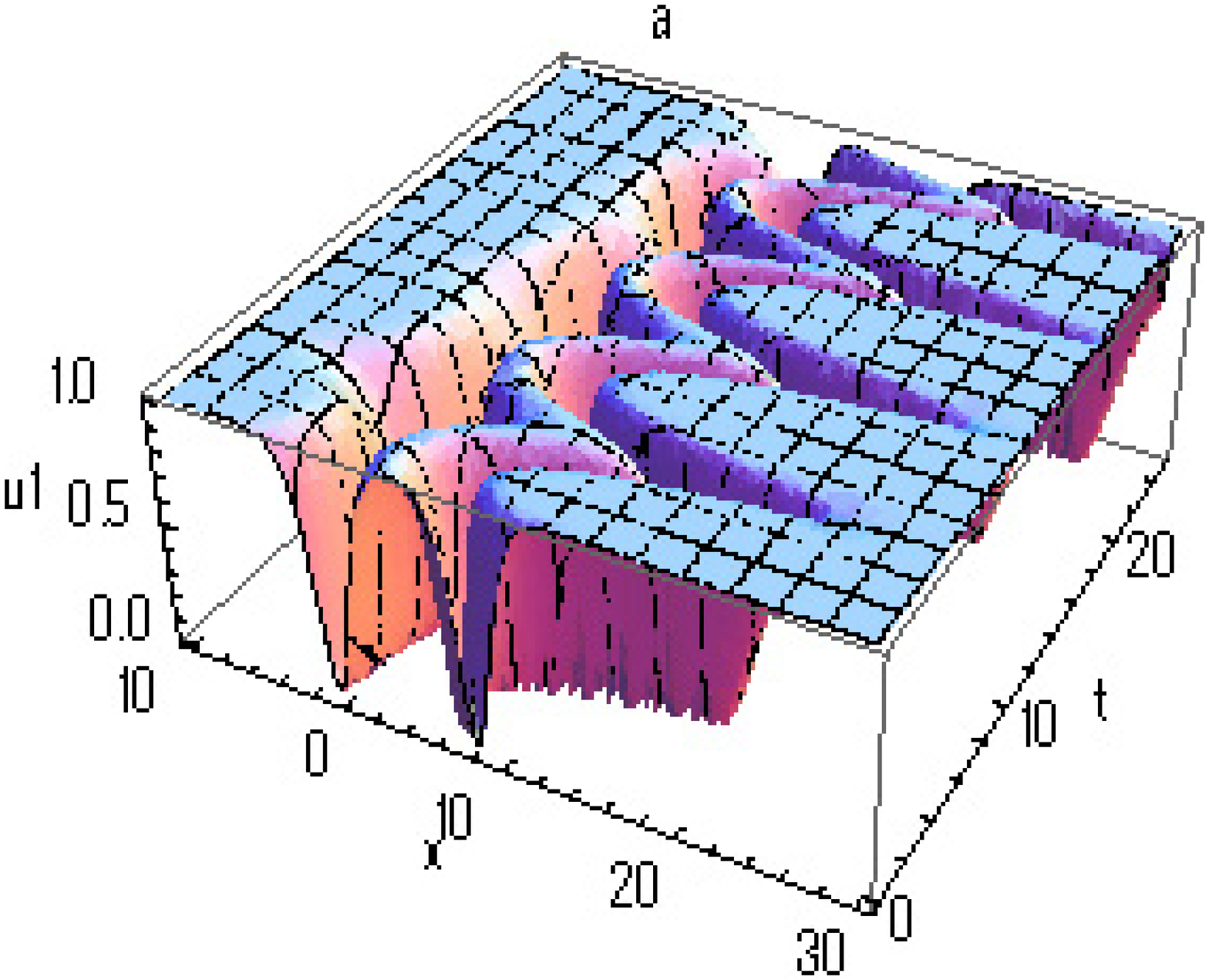}
\includegraphics[width=0.45\columnwidth]{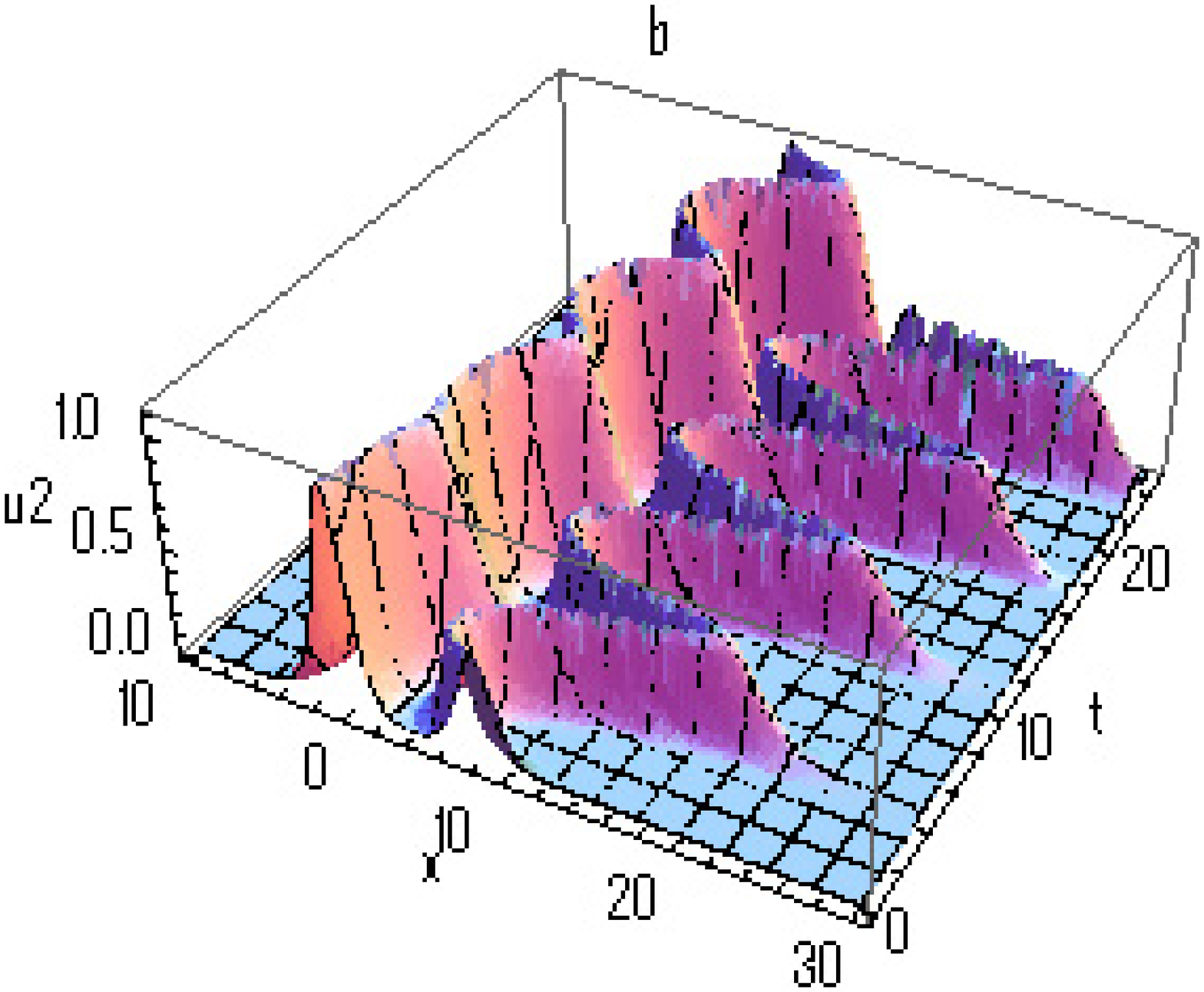}
\includegraphics[width=0.45\columnwidth]{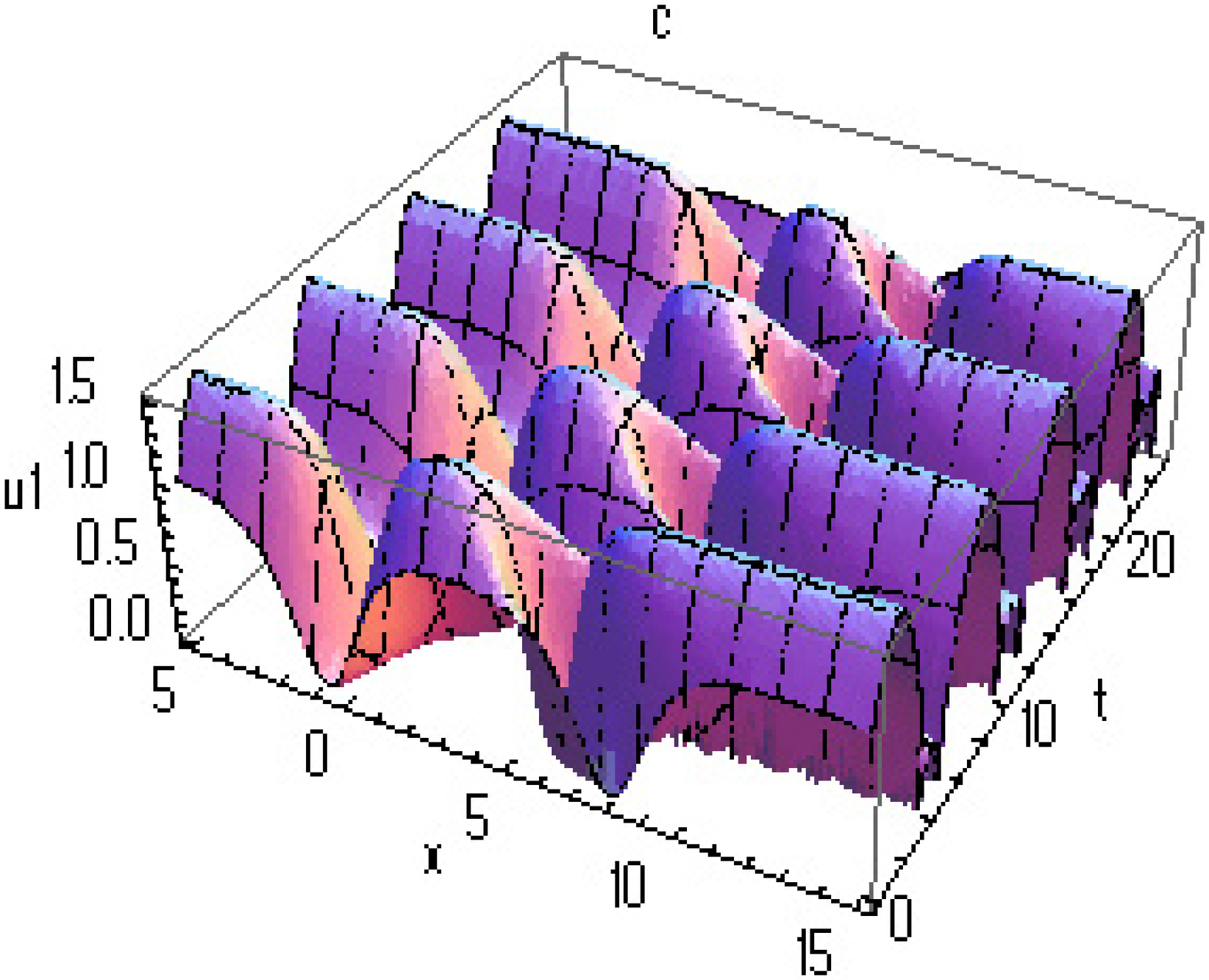}
\includegraphics[width=0.45\columnwidth]{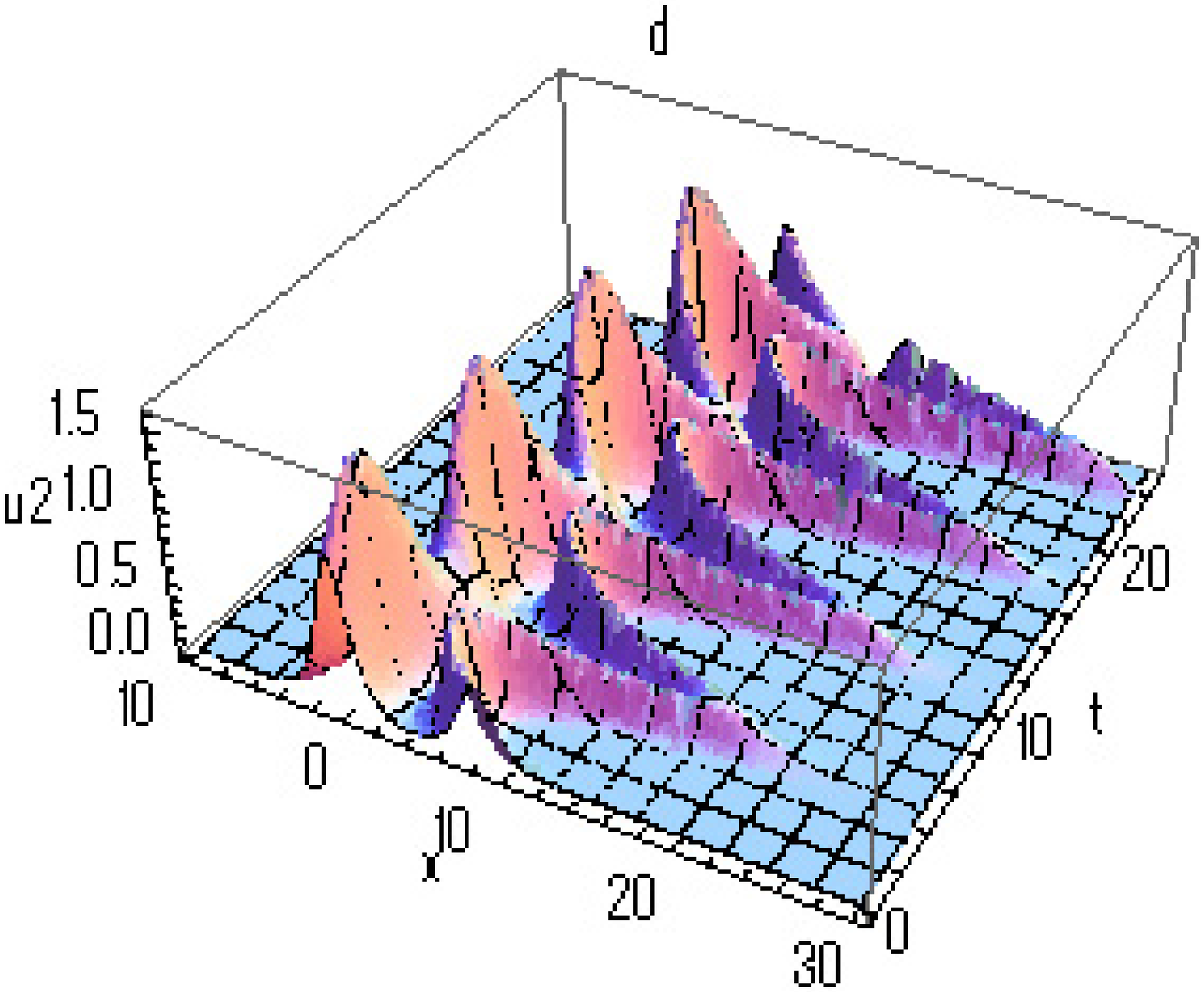}
\caption{(Color online) The standard two dark-bright solitons (a, b) and the
corresponding nonautonomous dark-bright solitons with the Feshbach
management (c, d). The parameters used are $\protect\lambda=0.5$ and $%
\protect\omega=1$. The other parameters used are $a_{1} = 1,$ $a_{2}=0.77,$ $%
b_{1} = b_{2} = 0,$ $c=0$ and $\protect\tau = 1. $}
\label{fig2}
\end{figure}

The generalized integrability condition of Eq. (\ref{eq6}) for the
two-component nonautonomous NLS equation has not been reported in the
literature. It can be used to investigate unique phenomenons in many
practical two-component nonlinear systems. To further explore it, first we
consider the symmetric case, such as $\varepsilon _{1}=\varepsilon
_{2}=\varepsilon$, $f_{1}(t)=f_{2}(t)$, $%
g_{11}(t)=g_{22}(t)=g_{12}(t)=g_{21}(t)=g(t)$ and $V_{1}(t)=V_{2}(t)=V(t)$.
Thus Eq. (\ref{eq6}) reduces to
\begin{equation}
2\varepsilon f(t)V(t)+\frac{d^{2}}{dt^{2}}\ln \frac{f(t)}{g(t)}-\frac{d}{dt}
\ln g(t)\frac{d}{dt}\ln \frac{f(t)}{g(t)}=0,  \label{eq8}
\end{equation}
which is the same as that in the single-component case. Furthermore, if $%
\varepsilon =1/2$ and $f(t)=1$, Eq. (\ref{eq8}) gives a constraint condition
on the applied external potential
\begin{equation}
V(t)=\frac{1}{g(t)}\frac{d^{2}}{dt^{2}}g(t)-\frac{2}{g(t)^{2}}\left( \frac{d
}{dt}g(t)\right) ^{2}.  \label{eq9}
\end{equation}
This expression is completely consistent with those reported in the
literature \cite{Zhang2009} by using different methods. Under this
condition, one may take a general transformation to obtain exact solution of
the coupled NLS equation (\ref{eq4}), as in the single-component case \cite%
{He2009}
\begin{equation}
u_{i}(x,t)=U_{i}(X(x,t),T(t))\mbox{e}^{ia(x,t)+c(t)},  \label{eq10}
\end{equation}%
where $U_{i}(i=1,2)$ satisfy the standard coupled NLS equation
\begin{eqnarray}
&&\hspace{-0.5cm}i\frac{\partial }{\partial T}U_{i}(X,T)+\varepsilon _{i}%
\frac{\partial ^{2}}{\partial X^{2}}U_{i}(X,T)+\delta
_{ii}|U_{i}(X,T)|^{2}U_{i}(X,T)  \notag \\
&&\hspace{3cm}+\delta _{ij}|U_{j}(X,T)|^{2}U_{i}(X,T)=0.
\label{eq11}
\end{eqnarray}
The explicit forms of $X(x,t),T(t),a(x,t)$ and $c(t)$ can be
obtained analytically as follows $ a(x,t)=\frac{1}{4\varepsilon
f(t)}\left( \frac{d}{dt}\ln \frac{f(t)}{g(t} \right)
x^{2}+C\frac{g(t)}{f(t)}x-\varepsilon C^{2}\int
\frac{g^{2}(t)}{f(t)} dt$, $X(x,t)=\frac{g(t)}{f(t)}x-2\varepsilon
C\int \frac{g^{2}(t)}{f(t)}dt$, $c(t)=\frac{1}{2}\ln \left(
\frac{g(t)}{f(t)}\right) $, and $T(t)=\int \frac{
g^{2}(t)}{f(t)}dt$, where $C$ is an arbitrary constant. Eq.
(\ref{eq11}) has a standard two-soliton solutions \cite{Adrian1997},
from which the corresponding nonautonomous two-soliton solutions of
Eq. (\ref{eq4}) can be explicitly obtained by Eq. (\ref{eq10}). To
understand the influence of Feshbach resonance on nonautonomous
two-dark-bright soliton behavior under the integrability condition
Eq. (\ref{eq8}), we investigate the dynamics of these solitons with
Feshbach management. These solitons can be realized in a
two-component system such as different hyperfine spin states
\cite{Myatt1997} . $g(t)$ is modulated periodically as $g(t) = 1 +
\lambda \sin( \omega t)$, where $\lambda$ and $\omega$ are the
relevant modulation parameters. The corresponding trap potential is
tuned by Eq. (\ref{eq9}). In Fig. \ref{fig2} we show the
nonautonomous two-dark-bright solitons dynamics obtained (c, d) and
also compare them with the canonical ones (a, b). The result shows
clearly that these solitons remain stable, as the conventional ones,
but their shapes and speeds change with time. It is significant to
guide experimental control over the dynamic behavior of the
muti-solitons in binary BECs.

\vspace{0.2cm}
\begin{figure}[h]
\includegraphics[width=0.45\columnwidth]{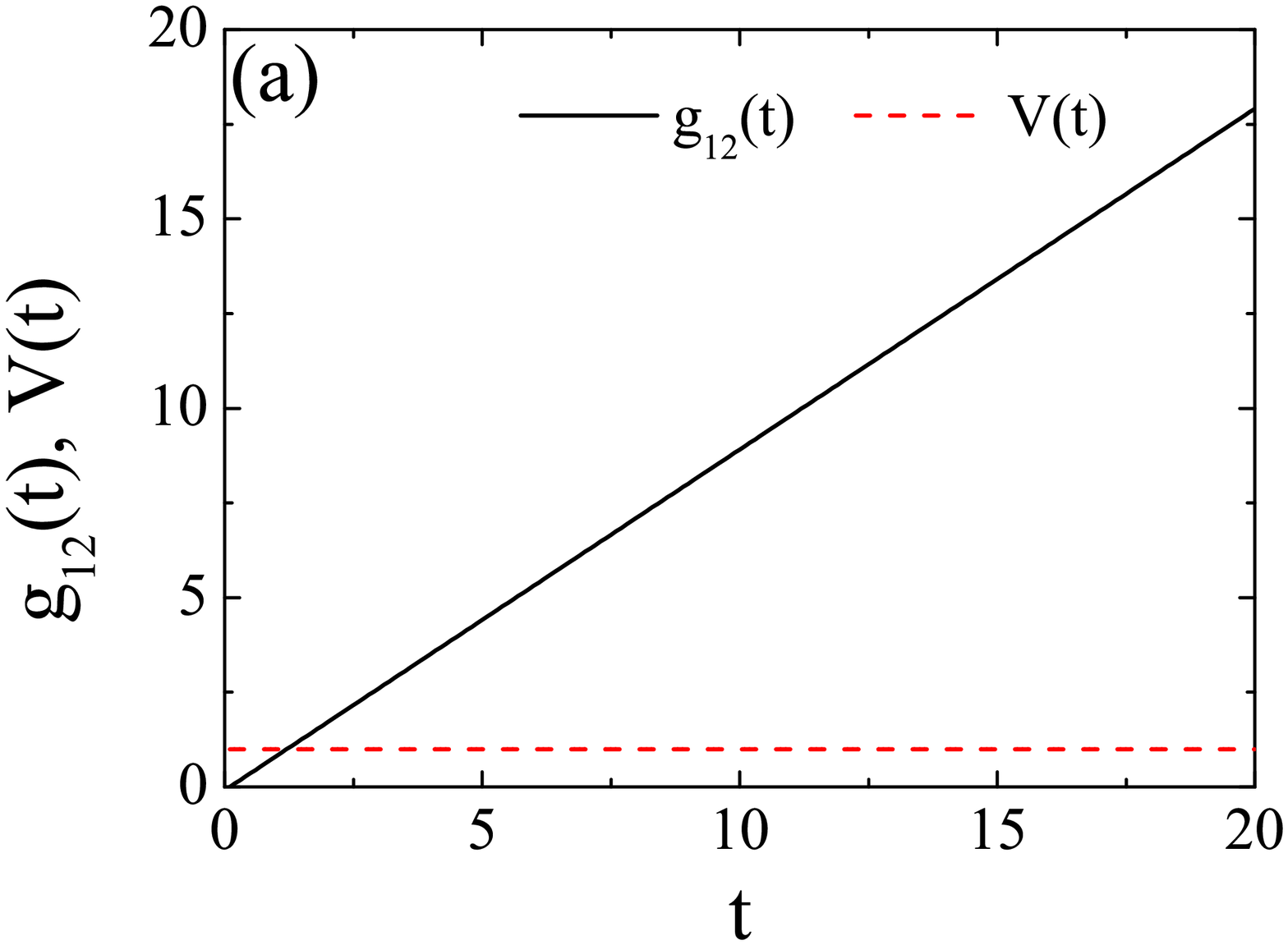} %
\includegraphics[width=0.45\columnwidth]{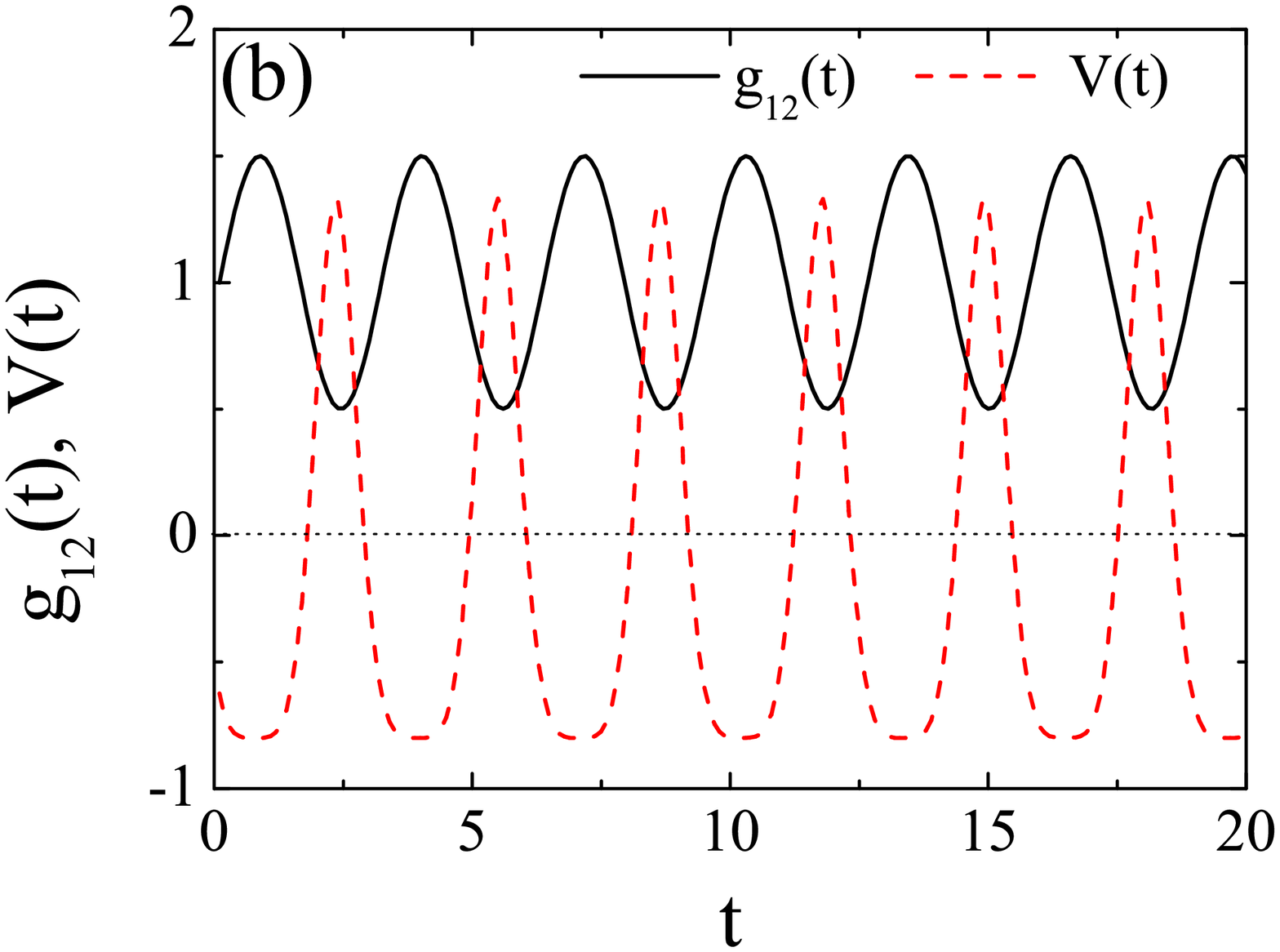} %
\includegraphics[width=0.45\columnwidth]{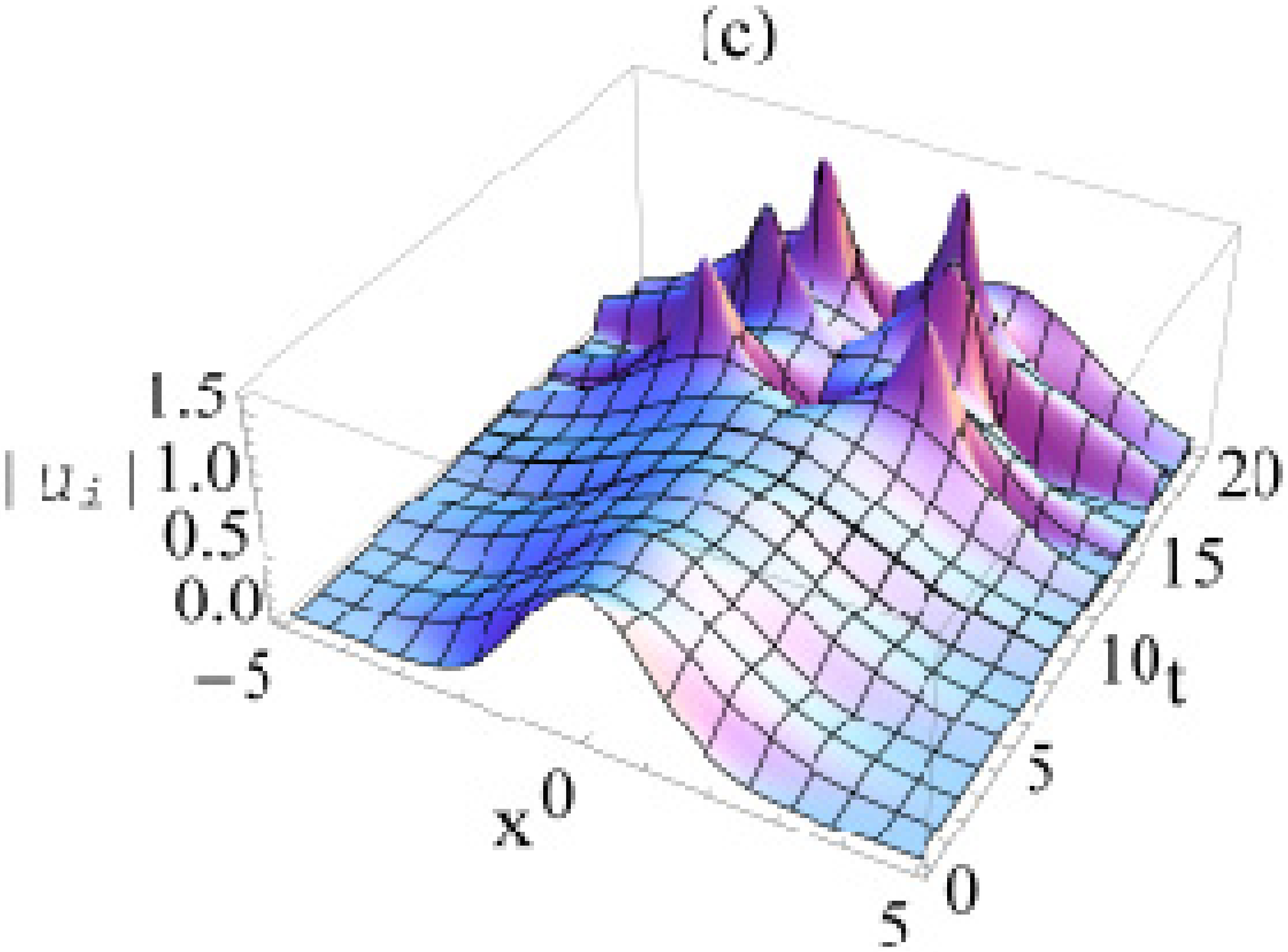} %
\includegraphics[width=0.45\columnwidth]{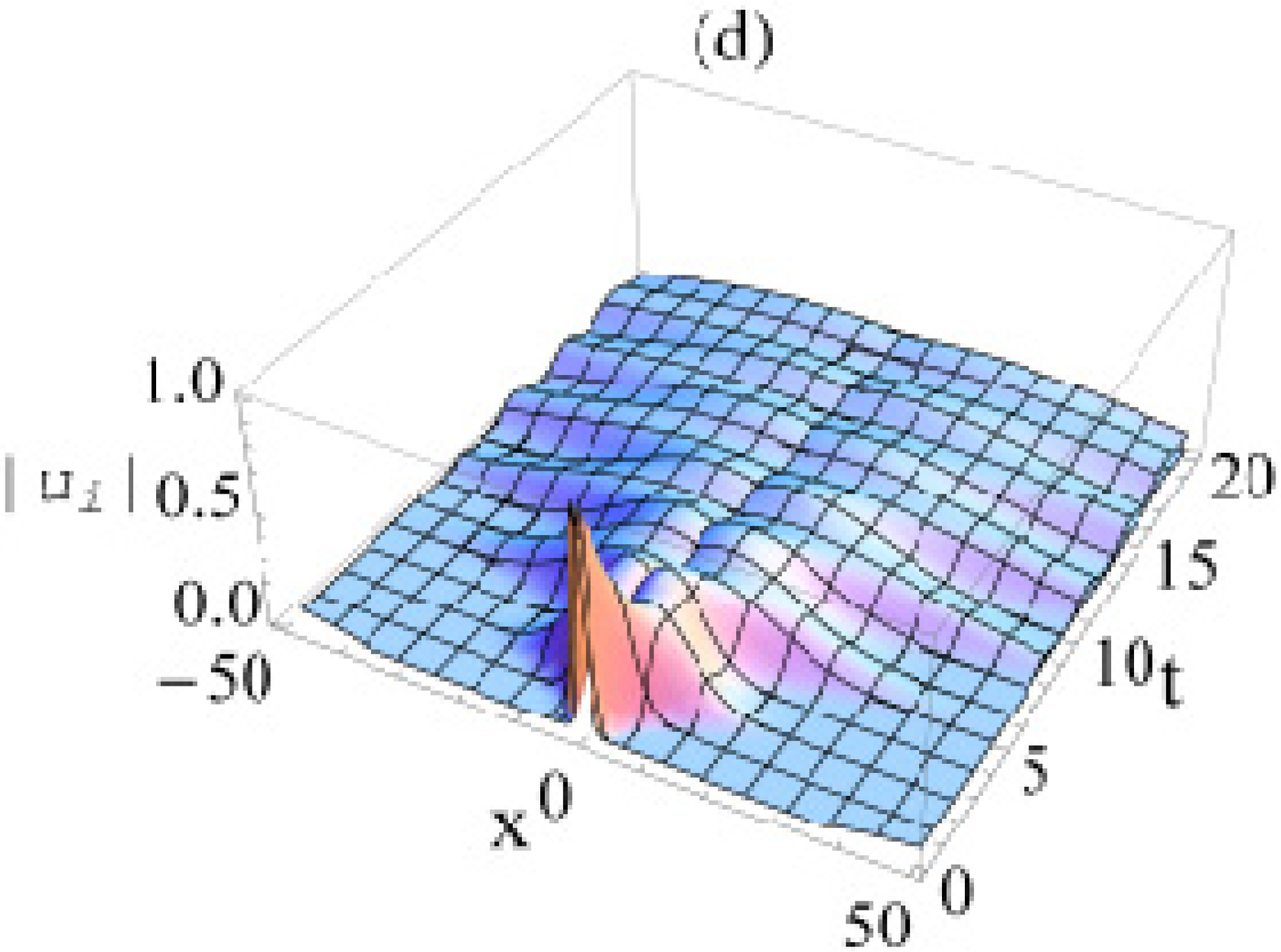}
\caption{(Color online) Spatial separation of two components in the
nonintegrable (a, c) case for $\protect\lambda=1$ and the integrable (b, d)
case for $\protect\lambda=0.5$ and $\protect\omega=2.$ }
\label{fig3}
\end{figure}
Finally, we consider the phase separation phenomenon in binary BEC. For
simplicity, we let $\delta _{11}=\delta _{22}=\delta _{12}=\delta _{21} = -1$%
, $f_{i}(t) =1$, $\varepsilon_{i}=1/2$, $g_{ii}(t) =1$ and $V_{i}(t) = V(t)$
($i = 1, 2$). Thus Eq. (\ref{eq6}) becomes
\begin{equation}
V(t) =\frac{1}{1+g_{12}(t)}\frac{d^{2}}{dt^{2}}g_{12}(t) -\frac{2}{(1 +
g_{12}(t)) ^{2}}\left(\frac{d}{dt}g_{12}(t)\right)^{2}.  \label{eq14}
\end{equation}
Unfortunately, the explicit solution of Eq. (\ref{eq4}) can not be obtained
directly from that of Eq. (\ref{eq11}) in this case, but it is enough to use
the Gaussian wave packets to explore the spatial separation problem. They
read \cite{Navarro2009}
\begin{eqnarray}
&& u_{1}(x,t) = A(t) e^{-\frac{\left( x-B(t) \right) ^{2}}{2W(t)^{2} }%
}e^{i\left(C(t) + D(t) x + E(t) x^{2}\right) },  \notag \\
&& u_{2}(x,t) = A(t) e^{-\frac{\left(x+B(t) \right) ^{2}}{2W(t)^{2} }%
}e^{i\left(C(t) - D(t) x+E(t) x^{2}\right)},  \label{eq15}
\end{eqnarray}
where $A(t),$ $B(t),$ $C(t),$ $D(t),$ $E(t)$ and $W(t)$ are the variational
parameters and are determined by a set of ordinary differential equations
given by Ref. \cite{Navarro2009}.

We solve this set of equations in two different cases. The first one is to
increase intercomponent interaction linearly but fix the harmonic external
potential as shown in Fig. \ref{fig3}(a). Initially the two components are
mixed and keep miscible for some time. Subsequently the spatial separation
happens. This is in agreement with the previous observation \cite%
{Navarro2009}. However, it is necessary to point out that this case
is not integrable. In contrast, in the second case we consider the
integrable constraint on the harmonic external potential given by
Eq. (\ref{eq14}). Here the intercomponent interaction assumed varies
around 1 periodically, as shown in Fig. \ref{fig3}(b). In this case
the applied external potential is modulated periodically following
the intercomponent interaction. The two components in this case are
found immiscible, even if the miscible condition $g_{12}(t) < 1$ is
satisfied frequently. This is a consequence of the change in the
external potential, or rather the presence of the potential barrier
($ V(t) < 0$) in some time, which leads to the two-component
separation shown in Fig. \ref{fig3}(d). The two different results
provide possible ways to realize two-component spatial separations
in binary BECs.

\section{Summary}

By analyzing the integrability conditions of the nonautonomous NLS
equations we found that the nature of the nonautonomous soliton is
due to the balance between different competition features: the
kinetic energy versus potential energy and the dispersion versus
nonlinearity. This picture is consistent with the conventional one
which originates from the balance between the dispersion and
nonlinearity. On the other hand, the integrability conditions in
two-component case is useful to guide experimental control over the
dynamic of nonautonomous solitons and two-component separation in
binary BECs.

\section{Acknowledgements}

The work is supported by the Program for NCET, NSF and the
Fundamental Research Funds for the Central Universities of China.
WML is supported by NSFC (No. 10934010) and the NKBRSFC (No.
2011CB921500).

\end{document}